\documentclass[]{interact}

\usepackage{epstopdf}
\usepackage{xcolor}
\usepackage[caption=false]{subfig}
\usepackage{physics}
\usepackage{textcomp}
\usepackage{longtable}
\usepackage{dcolumn}
\usepackage{xcolor}
\usepackage[normalem]{ulem}
\usepackage{amsmath}
\newcommand{\stkout}[1]{\ifmmode\text{\sout{\ensuremath{#1}}}\else\sout{#1}\fi}
\newcolumntype{d}[1]{D{.}{.}{#1}}
\usepackage[numbers,sort&compress,merge]{natbib}
\bibpunct[, ]{[}{]}{,}{n}{,}{,}
\theoremstyle{plain}

\theoremstyle{definition}

\theoremstyle{remark}

\begin{document}

\articletype{ARTICLE}

\title{Characterisation of the $b^3\Sigma^+, v=0$ State and Its Interaction with the $A^1\Pi$ State in Aluminium Monofluoride}%
%
\author{
\name{M.~Doppelbauer\textsuperscript{a}, N.~Walter\textsuperscript{a}, S.~Hofs\"{a}ss\textsuperscript{a}, S.~Marx\textsuperscript{a}, H.~C.~Schewe\textsuperscript{a}, S.~Kray\textsuperscript{a}, J.~P\'{e}rez-R\'{i}os\textsuperscript{a}, B.~G.~Sartakov\textsuperscript{b}, S.~Truppe \textsuperscript{a}\thanks{CONTACT S.~Truppe. Email: truppe@fhi-berlin.mpg.de}, G.~Meijer \textsuperscript{a}\thanks{CONTACT G. Meijer. Email: meijer@fhi-berlin.mpg.de}}
\affil{\textsuperscript{a}Fritz-Haber-Institut der Max-Planck-Gesellschaft, Faradayweg 4-6, 14195 Berlin, Germany \textsuperscript{b}General Physics Institute, Russian Academy of Sciences, Vavilovstreet 38, 119991 Moscow, Russia}
}
\maketitle

\begin{abstract}
Recently, we determined the detailed energy level structure of the $X^1\Sigma^+$, $A^1\Pi$ and $a^3\Pi$ states of AlF that are relevant to laser cooling and trapping experiments \cite{Truppe2019}. Here, we investigate the $b^3\Sigma^+, v=0$ state of the AlF molecule. A rotationally-resolved (1+2)-REMPI spectrum of the $b^3\Sigma^+, v'=0 \leftarrow a^3\Pi, v''=0$ band is presented and the lifetime of the $b^3\Sigma^+, v=0$ state is measured to be 190(2)~ns. Hyperfine-resolved, laser-induced fluorescence spectra of the $b^3\Sigma^+, v'=0 \leftarrow X^1\Sigma^+, v''=1$ and the $b^3\Sigma^+, v'=0 \leftarrow a^3\Pi, v''=0$ bands are recorded to determine fine- and hyperfine structure parameters. The interaction between the $b^3\Sigma^+, v=0$ and the nearby $A^1\Pi$ state is studied and the magnitude of the spin-orbit coupling between the two electronic states is derived using three independent methods to give a consistent value of 10(1)~cm$^{-1}$. The triplet character of the $A$ state causes an $A\rightarrow a$ loss from the main $A-X$ laser cooling cycle below the 10$^{-6}$ level.
\end{abstract}

\begin{keywords}
Cold molecules, laser cooling, hyperfine-resolved spectroscopy, spin-orbit coupling
\end{keywords}

\section{\label{sec:intro}Introduction}
Polar molecules, cooled to low temperatures, have wide-ranging applications in physics and chemistry \cite{Carr2009, Safronova2018}. Recently, we identified the AlF molecule as an excellent candidate for direct laser cooling to low temperatures and with a high density \cite{Truppe2019}. AlF has one of the strongest chemical bonds known (6.9 eV), can be produced efficiently, and captured and cooled in a magneto-optical trap using any Q-line of its strong $A^1\Pi, v'=0\leftarrow X^1\Sigma^+, v''=0$ band near 227.5~nm. Subsequent to trapping and cooling on the strong $A-X$ transition, the molecules may be cooled to a few \textmu K on any (narrow) Q-line of the spin-forbidden $a^3\Pi_1, v'=0\leftarrow X^1\Sigma^+, v''=0$ band. To laser-cool AlF successfully, it is essential to study the detailed energy level structure of the molecule, to measure the radiative lifetime of the states involved and to investigate and quantify the potential decay channels to states that are not coupled by the cooling laser, i.e. losses from the optical cycle.

For nearly a century, spectroscopists have been interested in the AlF molecule, in part due to its similarity to the much studied CO and N$_2$ molecules. AlF is also important to the astrophysics community and it has been detected in sunspots, stellar atmospheres, circumstellar envelopes and proto-planetary nebulae \cite{Highberger2001, Agundez2012}. Moreover, $^{26}$AlF was the first radioactive molecule to be discovered in space \cite{Kaminski2018}. AlF has been the subject of theoretical studies using \textit{ab initio} quantum chemistry to determine radiative lifetimes, dipole moments and potential energy curves for its electronic states \cite{So1974, Langhoff1988,Woon2009, Wells2011}. Precise spectroscopic parameters for AlF are useful for future astrophysical observations and new spectroscopic studies of electronic states can serve as a benchmark for quantum chemistry calculations.

In this paper, we present rotationally-resolved optical spectra of the $b^3\Sigma^+, v'=0\leftarrow a^3\Pi, v''=0$ band and demonstrate state-selective ionization of $a^3\Pi$ molecules via (1+2)-resonance-enhanced multi-photon ionization (REMPI). Pulsed excitation followed by delayed ionization with a KrF excimer laser is used to determine the radiative lifetime of the $b^3\Sigma^+, v=0$ state. Next, high-resolution, cw laser-induced fluorescence (LIF) spectra of the $b^3\Sigma^+, v'=0\leftarrow X^1\Sigma^+, v''=1$ band are presented. The hyperfine structure in the $b^3\Sigma^+, v=0$ state is resolved and precise spectroscopic constants for the $b$ state are determined. Laser-induced fluorescence spectra, recorded using a cw laser to drive rotational lines of the $b^3\Sigma^+, v'=0\leftarrow a^3\Pi, v''=0$ band, allow us to reduce the uncertainty in the spin-orbit constant $A$ and the spin-spin interaction constant $\lambda$ of the $a^3\Pi$ state by nearly two orders of magnitude. A fraction of the excited state molecules decays on the $b^3\Sigma^+, v'=0\rightarrow X^1\Sigma^+, v''$ bands to the ground state. This is caused by spin-orbit coupling of the $b^3\Sigma^+, v=0$ state with the nearby $A^1\Pi$ state. This perturbation of the $A^1\Pi$ state and the $b^3\Sigma^+$ state is analysed in detail as it induces also a small loss channel for the strong $A-X$ cooling transition. 

\section{\label{sec:history}Previous Work}
In 1953, Rowlinson and Barrow reported the first observation of transitions between triplet states in AlF by recording emission spectra from a hollow-cathode discharge \cite{Rowlinson1953}. In the following decades, new triplet states were identified and characterised in more detail \cite{Naude1953, Dodsworth1954, Dodsworth1955, Barrow1963, Kopp1970}. Kopp and Barrow analysed the interaction of the $A^1\Pi$ state with the nearby $b^3\Sigma^+$ state and thereby determined the term energy of the triplet states relative to the singlet states \cite{Kopp1970}. A comprehensive study of the electronic states and the interaction between the singlet and triplet states followed \cite{Barrow1974}. In 1976, Rosenwaks \textit{et al.} observed the $a^3\Pi\rightarrow X^1\Sigma^+$ transition directly in emission \cite{Rosenwaks1976} at the same time as Kopp  \textit{et al.} did in absorption \cite{Kopp1976}. Both studies confirmed the singlet-triplet separation determined from the earlier perturbation analysis.

The hyperfine structure of the rotational states in $b^3\Sigma^+$ was partly resolved and analysed by Barrow \textit{et al.} \cite{Barrow1974}. Their low-resolution spectra (typical linewidth of 0.05~cm$^{-1}$) showed a triplet structure, which they ascribed to magnetic hyperfine effects of the $^{27}$Al nucleus, whose nuclear spin is I$_\textrm{Al}=5/2$. A more detailed analysis of both the fine and hyperfine structure in the triplet states was presented by Brown \textit{et al.} in 1978 \cite{Brown1978}. The hyperfine structure was partially resolved and they determined values for the Fermi contact parameter $b_F(\textrm{Al})$, the electron spin-spin interaction parameter $\lambda$, and the spin-rotation parameter $\gamma$.

We have recently reported on the detailed energy level structure of the $X^1\Sigma^+$, $A^1\Pi$ and $a^3\Pi$ states in AlF \cite{Truppe2019}. That study resolves the complete hyperfine structure, including the contribution due to fluorine nuclear spin, in all three states. The energy levels in $X^1\Sigma^+$ and in the three $\Omega$ manifolds of the metastable $a^3\Pi$ state were measured with kHz resolution; this allowed us to determine the spectroscopic constants precisely and study subtle effects, such as a spin-orbit correction to the nuclear quadrupole interaction of the Al nucleus.\\

\section{\label{section:overview}Rotational Structure of the $b^3\Sigma^+, v'=0\leftarrow a^3 \Pi_1, v''=0$ Band and the Radiative Lifetime of the $b^3\Sigma^+, v=0$ State}

To study the $b-a$ transition, the molecules must first be prepared in the $a^3\Pi$ state. Previously, we described this state and the $a-X$ transition in detail \cite{Truppe2019}. The hyperfine structure in $X^1\Sigma^+$ is small and for the purpose of this study, the ground state is comprised of single rotational levels with energies given by $B J(J+1)$.

The angular momentum coupling in the $a^3\Pi$ state of AlF is well-described by Hund's case (a). The energy levels are labelled by the total angular momentum (excluding hyperfine interaction) $\mathbf{J}=\mathbf{R}+\mathbf{L}+\mathbf{S}$, with $\mathbf{R}$ the rotational angular momentum of the rigid nuclear framework, and $\mathbf{L}$ and $\mathbf{S}$ the total orbital and spin angular momenta of the electrons, respectively. Both, $\mathbf{L}$ and $\mathbf{S}$ are not well defined. However, their projection onto the internuclear axis $\mathbf{\Lambda}$, $\mathbf{\Sigma}$ and the total electronic angular momentum along the internuclear axis $\mathbf{\Omega}=\mathbf{\Lambda}+\mathbf{\Sigma}$ are well defined. The spin-orbit interaction leads to three fine-structure states with $\Omega=|\Lambda+\Sigma|$, labelled with  $\mathcal{F}_1$, $\mathcal{F}_2$ and $\mathcal{F}_3$, in order of increasing energy, corresponding to the states $a^3\Pi_0$, $a^3\Pi_1$ and $a^3\Pi_2$. The splitting between the $\Omega$ manifolds is determined by $A$, the electron spin-orbit coupling constant and the spin-spin interaction coupling constant $\lambda$. Each $\Omega$ has its own rotational manifold with energies $B J(J+1)$, where the rotational constant $B$ is slightly different for each $\Omega$ manifold. 

The $b^3\Sigma^+$ state is well-described by Hund's case (b), for which $\mathbf{L}$ couples to the rotation $\mathbf{R}$ to form $\mathbf{N}=\mathbf{R}+\mathbf{L}$. The rotational energy levels follow $BN(N+1)$. For $\Sigma$ electronic states $\mathbf{N}=\mathbf{R}$ and the total angular momentum (without hyperfine interaction) is $\mathbf{J}=\mathbf{N}+\mathbf{S}$. The combined effect of the spin-rotation interaction $\gamma(\mathbf{N}\cdot\mathbf{S})$ and the spin-spin interaction $\frac{2}{3}\lambda(3S^2_z-\mathbf{S}^2)$ splits each rotational level $N>0$ of $b^3\Sigma^+$ into three J-levels; these components are labelled $\mathcal{F}_1$, $\mathcal{F}_2$ and $\mathcal{F}_3$ with quantum numbers $J=N+1$, $J=N$ and $J=N-1$, respectively. 

Following the convention of Brown \textit{et al.} \cite{Brown1975} the parity states can be labelled by $e$ and $f$, where $e$ levels have parity $+(-1)^J$ and $f$ labels have parity $-(-1)^J$. All rotational levels of the $X^1\Sigma^+$ state are $e$-levels, while in the $b^3\Sigma^+$ state, all $\mathcal{F}_2$ levels are $e$-levels, while $\mathcal{F}_1$ and $\mathcal{F}_3$ are $f$-levels. In the $a^3\Pi$ state, each J-level has an $e$ and an $f$ component due to $\Lambda$-doubling. 

\begin{figure}[htb!]
\centering
\includegraphics[width=\linewidth]{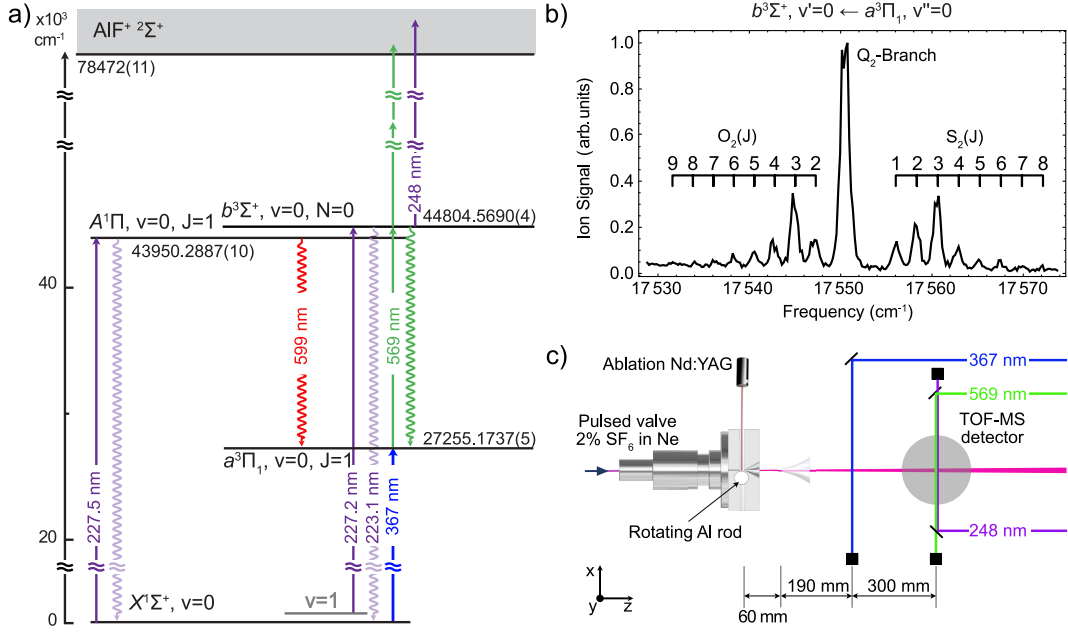}
\caption{\label{fig:Figure_1} a) Electronic energy level scheme of the relevant states of AlF. The transitions used for laser excitation are shown as solid arrows. The laser-induced fluorescence used for detecting the molecules is indicated by downward wavy arrows. The indicated energies are the gravity centres of the respective states in absence of hyperfine structure. b) (1+2)-REMPI spectrum of the $b^3\Sigma^+, v'=0\leftarrow a^3\Pi_1, v''=0$ band. The $a^3\Pi_1, v=0$ state is populated via the Q-branch of the $a^3\Pi_1, v'=0\leftarrow X^1\Sigma^+, v''=0$ band using a frequency-doubled pulsed dye laser. c) Schematic of the experimental setup used for the determination of the lifetime of the $b^3\Sigma^+, v=0$ state.}
\end{figure}

Figure \ref{fig:Figure_1}a shows the energy level diagram of the electronic states relevant to this study, together with a rotationally resolved spectrum of the $b^3\Sigma^+, v'=0 \leftarrow a^3\Pi_1, v''=0$ band in \ref{fig:Figure_1}b, and a sketch of the experimental setup in \ref{fig:Figure_1}c, which is similar to the one reported previously \cite{Truppe2019}. The molecules are produced by laser-ablating an aluminium rod in a supersonic expansion of 2\%~SF$_6$ seeded in Ne. After passing through a skimmer, the ground-state molecules are optically pumped to the metastable $a^3\Pi_1, v=0$ state by a frequency-doubled pulsed dye laser using the Q-branch of the $a^3\Pi_1, v'=0 \leftarrow X^1\Sigma^+, v''=0$ band. For this, 367~nm radiation with a bandwidth of 0.1~cm$^{-1}$ and a pulse energy of 6~mJ in a beam with a $e^{-2}$ waist radius of about 2~mm is used. The Q-branch of this transition falls within this bandwidth. Therefore, many rotational levels in the metastable $a^3\Pi_1, v=0$ state are populated simultaneously. Via the Q-branch, only the $f$-levels in $a^3\Pi_1$ are populated. Alternatively, if the molecules are optically pumped to the metastable state using spectrally isolated R or P lines, only the $e$-levels are populated. Further downstream, at $z=55$~cm from the source, the molecular beam is intersected with light from a second pulsed dye laser tuned to the $b^3\Sigma^+, v'=0 \leftarrow a^3\Pi_1, v''=0$ transition near 569~nm. For pulse energies exceeding 6~mJ (unfocused, with an $e^{-2}$ waist radius of about 5~mm), this laser transfers population to the $b^3\Sigma^+, v=0$ state and subsequently ionises the molecules by having them absorb two more photons from the same laser. Such a one-colour (1+2)-REMPI scheme using an unfocused laser beam is very uncommon. However, AlF has numerous electronically excited states that lie one photon-energy above the $b$ state energy, strongly enhancing the non-resonant two-photon ionisation probability. The ions are mass-selected in a short time-of-flight mass spectrometer (TOF-MS) and detected using micro-channel plates. The TOF-MS voltages are switched on shortly after the ionisation laser fires; this way ionisation occurs under field-free conditions and the states have a well-defined parity. The (1+2)-REMPI scheme uses low-energy photons and an unfocused laser beam, which has the benefit of producing a mass spectrum with only a single peak, corresponding to AlF. The spectrum, displayed in Figure \ref{fig:Figure_1}b, shows the ion signal as a function of the REMPI laser frequency. The spectral lines are labelled by $\Delta(NJ)_{\mathcal{F}''}(J'')$, where $\Delta(NJ)=N'-J''$, as the quantum number $J$ is not well-defined in the $b$ state, {\textit{vide infra}}. Since only the $f$-levels of the $a^3\Pi$ state are populated, the spectrum consists of $\Delta(NJ)=-2,0,+2$, i.e., O, Q and S branches.

\begin{figure}[htb]
\centering
\includegraphics{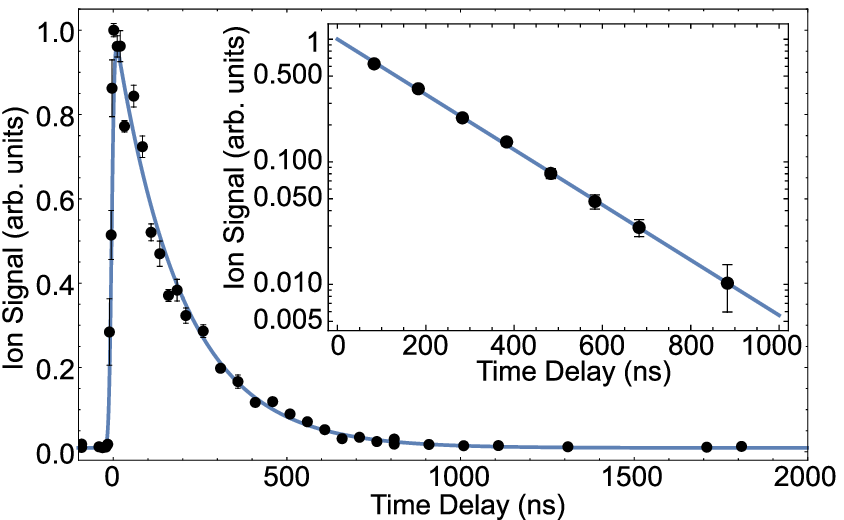}
\caption{\label{fig:Figure_2} Measurement of the radiative lifetime of the $b^3\Sigma^+, v=0$ state. First, the ground-state molecules are optically pumped to the $b^3\Sigma^+, v=0$ state via the two-colour excitation scheme described in the text. The population in the $b^3\Sigma^+, v=0$ state is probed by single-photon ionisation using a KrF excimer laser, followed by TOF-MS detection. The black dots show the AlF$^+$ ion signal as a function of the time delay between excitation and ionisation. The blue line is a fit to the data using the model described in the text. The inset shows a semi-log plot of a second measurement for eight specific time delays. A linear fit to the data gives a $b$ state lifetime of 190(2)~ns.}
\end{figure}

To determine the radiative lifetime of the $b^3\Sigma^+, v=0$ state, we reduce the pulse energy of the $b\leftarrow a$ excitation laser to about 1~mJ. At pulse energies below 2~mJ, the $a^3\Pi_1$ molecules are excited to the $b$ state, but are not ionised via (1+2)-REMPI. Instead, a KrF excimer laser is used to ionise the $b$ state molecules with a single 248~nm photon and a pulse energy of about 3~mJ. The radiative lifetime of the $b^3\Sigma^+, v=0$ state is measured by varying the time-delay between excitation and ionisation \cite{Duncan1981}. The determination of radiative lifetimes in the range of 20~ns to 1~\textmu s is relatively straightforward, because it is longer than the laser pulse duration, but short enough so that the molecules do not leave the detection region. Figure \ref{fig:Figure_2} shows the ion signal (black dots) as a function of the time delay with $1-\sigma$ standard error bars. The blue line is a fit to the data using the model $S(t)=C e^{-t/\tau_b}\textrm{erfc}[(t_0-t)/(\sqrt{2}\sigma)]$, where $\textrm{erfc}$ is the complementary error function, $C$ and $t_0$ are fit parameters, $\tau_b$ is the lifetime of the $b^3\Sigma^+, v=0$ state and $\sigma$ is the measured, combined pulse-width of the excitation and ionisation laser. In this model the lifetime $\tau_b=190(2)$~ns is fixed to the value determined from a linear fit to the semi-log plot shown in the inset. In 1988, Langhoff et al. calculated an approximate radiative lifetime of the $b^3\Sigma^+$ state of 135~ns \cite{Langhoff1988}, considerably shorter than the measured lifetime.\\

\section{\label{section:b3Shyperfine}The Fine and Hyperfine Structure of the $b^3\Sigma^+$ State}
Aluminium and fluorine have a nuclear spin of $I_{\textrm{Al}}=5/2$ and $I_{\textrm{F}}=1/2$, respectively. Following the description of Brown et al. \cite{Brown1978} and including the magnetic interaction of the fluorine nuclear spin, the effective Hamiltonian reads

\begin{align}
    H_{\textrm{eff}}&= \frac{2}{3}\lambda\left(3S_z^2-\mathbf{S}^2\right)+\gamma\left(\mathbf{N}\cdot\mathbf{S}\right)\\
    &+b_F(\textrm{Al}) \mathbf{I_{\textrm{\textrm{Al}}}}\cdot\mathbf{S}+\frac{1}{3}c(\textrm{Al})(3I_{\textrm{Al},z}S_z-\mathbf{I}_{\textrm{Al}}\cdot\mathbf{S})\nonumber\\
    &+b_F(\textrm{F}) \mathbf{I_{\textrm{F}}}\cdot\mathbf{S}+\frac{1}{3}c(\textrm{F})(3I_{\textrm{F},z}S_z-\mathbf{I}_{\textrm{F}}\cdot\mathbf{S})\nonumber\\
    &+\frac{eq_0Q}{4I_{\textrm{Al}}(2I_{\textrm{Al}}-1)}(3I^2_{\textrm{Al},z}-I^2_{\textrm{Al}})\nonumber,
\end{align}

where $\lambda$ is the spin-spin interaction constant and $\gamma$ the spin-rotation interaction constant. The parameters $b_F(\textrm{Al})$ and $b_F(\textrm{F})$ describe the Fermi contact interaction for the aluminium and fluorine nucleus, respectively, $c(\textrm{Al})$ and $c(\textrm{F})$ the dipolar interaction, and $eq_0Q$ the electric quadrupole interaction of the Al nucleus.\\

\begin{figure}[!htb]
\centering
\includegraphics[width=\linewidth]{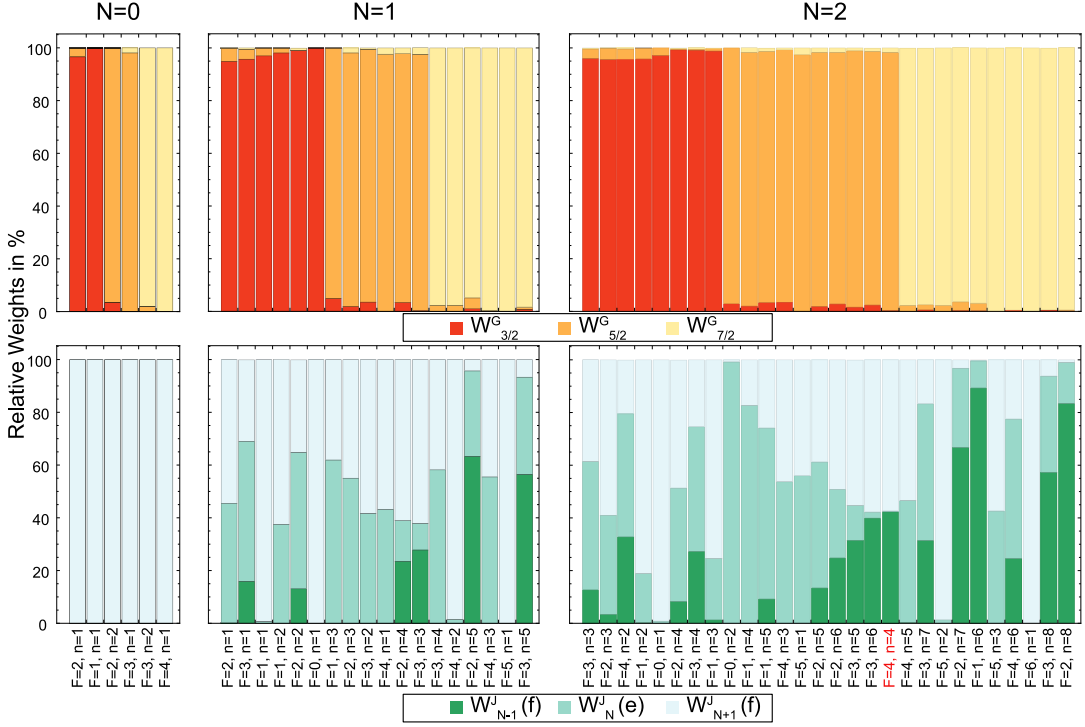}
\caption{\label{fig:Figure_4} Relative weights $W^{G}$ and $W^J$ of the $\ket{N,G,F}$ and $\ket{N,J,F}$ basis wave functions in the eigenfunctions $\ket{E}$ of the energy levels in the three lowest rotational states of the $b^3\Sigma^+, v=0$ state. The weights are obtained by projecting the eigenstate of the Hamiltonian onto the basis wave function $\ket{N,G,F}$ and $\ket{N,J,F}$, respectively. $G$ can take the values $3/2, 5/2$ and $7/2$ and $J$ can be $N-1, N$ or $N+1$. The energy levels are described uniquely by the quantum numbers $F$, $p$ and the index $n$. This analysis shows that $G$ is a good quantum number for the description of any rotational level of the $b^3\Sigma^+, v=0$ state, while $J$ only characterises the levels well for $N=0$.}
\end{figure}

In the $b^3\Sigma^+$ state of AlF the Fermi contact interaction $b_F(\textrm{Al})\left(\mathbf{I}_{\textrm{Al}}\cdot \mathbf{S}\right)$ between the nuclear spin of aluminium and the electronic spin angular momentum is strong compared to the spin-rotation interaction \cite{Brown1978}. The coupling case approximates $(b_{\beta S})$. The $N=0$ level has only one spin-component for which $J=N+S=1$; this $J=1$ level is split into three components due to the aluminium nuclear spin, each of which is again split by the nuclear spin of fluorine. This results in a total of six $F$ levels. For $N>0$, $J$ is not well defined and it is useful to introduce an intermediate angular momentum $\mathbf{G}=\mathbf{I}_{\textrm{Al}}+\mathbf{S}$ (see Figure \ref{fig:Figure_4}). $\mathbf{G}$ then couples to $\mathbf{N}$ which results in three sets of sub-levels, with quantum numbers $G=3/2,5/2$ and 7/2, each of which contains the closely spaced stacks of $N+3/2,...N-3/2$,  $N+5/2,...,N-5/2$ and $N+7/2, N+5/2...,N-7/2$ levels. Each of these levels is again split by the interaction with the fluorine nuclear spin to give the total angular momentum $\mathbf{F}=\mathbf{N}+\mathbf{G}+\mathbf{I_{\textrm{F}}}$. For $N=1,2$ and 3 this results in a total of 18, 28 and 34 $F$ levels, respectively. For $N>3$ the number of $F$ levels reaches its limit of 36.

To determine the hyperfine energy levels of the $b^3\Sigma^+$ state, we drive the $b^3\Sigma^+, v'=0 \leftarrow X^1\Sigma^+, v''=1$ band near 227.2~nm with a cw laser and detect the $b-a$ fluorescence at 569~nm. The spectrum of this transition directly reflects the energy level structure in the $b^3\Sigma^+$ state, because the hyperfine structure of the $X^1\Sigma^+$ state is smaller than the residual Doppler broadening in the molecular beam.  The wavelength required to drive this transition is close to that of the $A^1\Pi, v'=0-X^1\Sigma, v''=0$ band near 227.5~nm, for which we have a powerful UV laser system installed. 

\begin{figure}[htb!]
\centering
\includegraphics{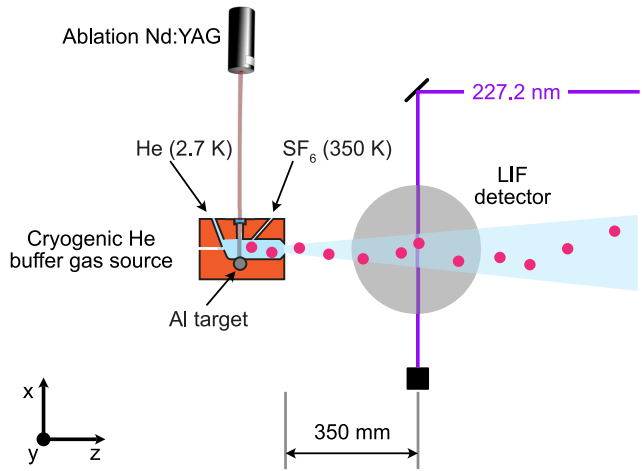}
\caption{\label{fig:Figure_3} Schematic of the experimental setup used to measure the fine and hyperfine structure of the $b^3\Sigma^+, v=0$ state. AlF molecules are produced in a cryogenic helium buffer gas source. The molecular beam is intersected with UV laser light from a frequency-quadrupled cw titanium sapphire laser. The molecules are excited on the weak, spin-forbidden $b^3\Sigma^+, v'=0\leftarrow X^1\Sigma^+, v''=1$ transition. The laser-induced fluorescence occurs mainly on the $b^3\Sigma^+, v'=0 \rightarrow \textrm{a}^3\Pi, v''=0$ transition near 569~nm and is imaged onto a PMT.}
\end{figure}

The $b^3\Sigma^+, v'=0 \leftarrow X^1\Sigma^+, v''=1$ transition is spin-forbidden and the overlap of the vibrational wave functions is poor, with a calculated Franck-Condon factor of 0.02. However, the transition becomes weakly allowed due to the spin-orbit interaction between the $b^3\Sigma^+$ and the nearby $A^1\Pi$ state. In section \ref{section:perturbations} this interaction will be discussed in more detail. To compensate for the weak transition dipole moment, we use a high excitation laser intensity of up to 75~mW in a laser beam with a $e^{-2}$ waist radius of 0.6 mm. The resulting laser-induced fluorescence occurs mainly on the dipole-allowed $b-a$ transition near 569~nm. The far off-resonant fluorescence allows us to block scattered laser light with a bandpass filter and to record background-free spectra. To increase the number of molecules in $X^1\Sigma^+, v=1$, we use a cryogenic helium buffer gas source, instead of the supersonic molecular beam introduced in the previous section. Figure \ref{fig:Figure_3} shows a sketch of the experimental setup that is used for this measurement.

The design of this source is similar to the one described in \cite{Maxwell2005,Hutzler2012, Truppe2018}. A pulsed Nd:YAG laser ablates a solid aluminium target in the presence of a continuous flow of 0.01~sccm room-temperature SF$_6$ gas, which is mixed with 1~sccm of cryogenic helium gas (2.7~K) inside a buffer gas cell. The hot Al atoms react with the SF$_6$ and form hot AlF molecules, which are subsequently cooled through collisions with the cold helium atoms. Compared to the supersonic molecular beam, this source delivers over 100 times more AlF molecules per pulse in the ro-vibronic ground-state to the detection region. The forward velocity of the molecules is four times lower. To increase the number of molecules in $X^1\Sigma^+, v=1$, we increase the pulse energy and repetition rate of the ablation laser, increase the temperature of the SF$_6$ gas to 350 K and increase its mass flow rate to 0.1~sccm. At $z=35$~cm the molecules interact with cw UV laser light to drive the $b^3\Sigma^+, v'=0 \leftarrow X^1\Sigma^+, v''=1$ transition near 227.2~nm. The laser light is produced by frequency-doubling the output of a cw titanium sapphire laser twice. The laser-induced fluorescence passes through an optical filter to block scattered light from the excitation laser, and is imaged onto a photomultiplier tube (PMT). The photo-current is amplified and acquired by a computer. The wavelength of the excitation laser is recorded with an absolute accuracy of 120~MHz using a calibrated wavemeter (HighFinesse WS8-10). 

\begin{figure}[htb!]
\centering
\includegraphics[width=\linewidth]{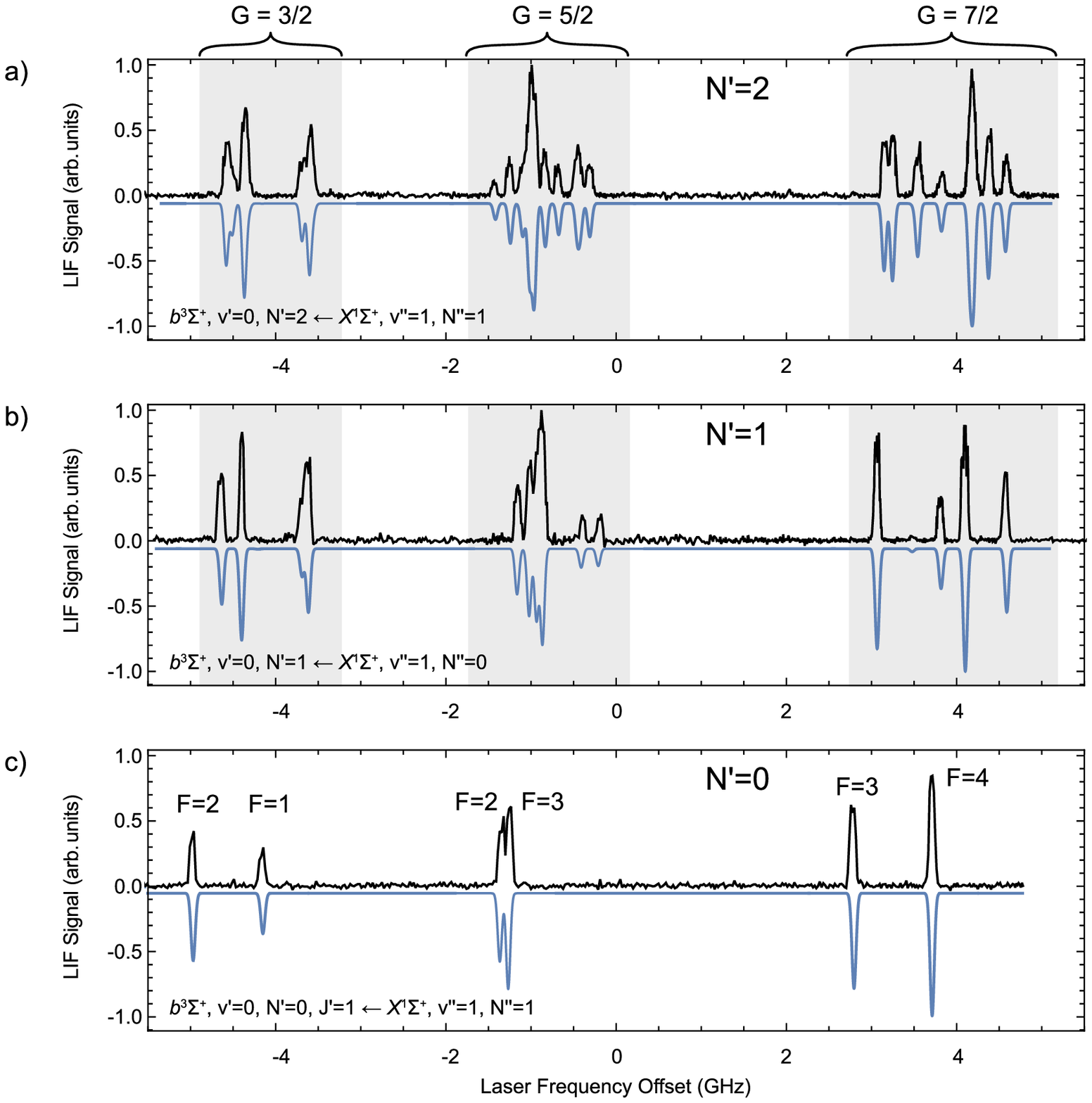}
\caption{\label{fig:Figure_5} Direct measurement of the fine and hyperfine structure of the three lowest rotational levels in the $b$ state, i.e. for $N'=0$ in panel c), $N'=1$, in panel b) and $N'=2$ in panel a). The experimental data are shown in black, pointing up and the simulated spectra are shown in blue, pointing down. The spectra are centred at the gravity centre, i.e., the line position in absence of hyperfine, spin-spin and spin-rotation interaction.}
\end{figure}

Figure \ref{fig:Figure_5} shows the recorded spectra reaching the three lowest $N$ levels in the $b$ state. The panels demonstrate the increasing complexity of the hyperfine structure with increasing $N$. The three spectra allow us to measure the rotational constant of the $b$ state. Gaussian lineshapes are fitted to the experimental spectra to determine the line-centres. We then fit the eigenvalues of the hyperfine Hamiltonian to the measured energy levels with the spectroscopic parameters as fit parameters. We assign a total of 48 lines and the standard deviation of the fit is 11 MHz. The best fit parameters together with their standard deviations are summarised in Table \ref{table:Table_1}. $E_0$ is the pure vibronic energy of $b^3\Sigma^+, v=0$, i.e. the energy of the $N=0$ level in absence of spin, fine and hyperfine splitting. This is referenced to the $J=0$ level of the $X^1\Sigma^+, v=0$ state by using the precise infrared emission lines of \cite{Hedderich1992} to determine the energy difference between the $v=0$ and $v=1$ level in the $X$ state. The inverted spectra in Figure \ref{fig:Figure_5} are simulated spectra using the spectroscopic parameters presented in Table \ref{table:Table_1} and reproduce the measured spectra well. The Fermi contact parameter $b_F(\textrm{F})$ for fluorine and the two hyperfine parameters $c(\textrm{Al})$ and $c(\textrm{F})$ for the aluminium and fluorine nucleus, respectively, are determined for the first time. In previous, low-resolution studies, the interaction of the fluorine nuclear spin has been neglected. However, we conclude that the magnitude of the interaction parameter for the two nuclei is comparable. This indicates that there is a significant electron density at both nuclei which is in stark contrast to the situation in the $a^3\Pi$ state. In the latter state, the Fermi contact term for the $^{19}$F nucleus is about seven times smaller than for the $^{27}$Al nucleus \cite{Truppe2019}. The Fermi contact parameter for the aluminium nucleus $b_F(\textrm{Al})$ and the spin-spin interaction parameter $\lambda$ presented here are consistent with the previously determined values, but their uncertainty is reduced significantly. 

\begin{table}[htb!]
\centering
\caption{\label{table:Table_1}Experimentally determined spectroscopic constants of $b^3\Sigma^+, v=0$. $E_0$  and its uncertainty is given in cm$^{-1}$, all other parameters are given in MHz. SD gives the standard deviation in the absence of correlations and $\textrm{SD}\cdot\sqrt{Q}$ gives the standard deviation of the parameter including the correlations between the parameters \cite{Watson1977}. The previous best values are taken from \cite{Barrow1974} and \cite{Brown1978} and given with the reported standard error (SE).}
 \begin{tabular}{cccccc} \hline\hline
 Parameter & This Work & SD & SD$\cdot\sqrt{Q}$& Refs. \cite{Barrow1974} \& \cite{Brown1978}& SE\\
 \hline
 $E_0$                  &     44804.5692    &       0.0004  &       0.0004 &   &\\
 $B$                    &      16772        &          2    &          5   &   16774.6 &   0.2\\
 $\lambda$              &       -919        &         15    &         18   &   -750    &   300\\
 $\gamma$               &         -9        &          7    &         13   &   0   &   9\\
 $b_F(\mbox{Al})$       &       1311        &          2    &          3   &   1469    &   90\\
 $c(\mbox{Al})$         &         73        &         12    &         18   &       &   \\
 $eq_0Q(\mbox{Al})$     &        -62        &         69    &         99   &       &\\
 $b_F(\mbox{F})$        &        870        &         10    &         11   &       &\\
 $c(\mbox{F})$          &        305        &         50    &         53   &       &\\
\hline\hline
\end{tabular}
\end{table}

\begin{table}
\centering
\caption{\label{table:Table_2} Energies, $E$, of the hyperfine levels in $b^3\Sigma^+, v=0$, relative to the $X^1\Sigma^+, v=0, J=0$ level, magnetic g-factors $g_F$, rotational quantum number $N$, total angular momentum $F$ and parity $p$. Since the assignment of the quantum numbers $N, F, p$, is not unique, we use $n$ to index the levels that share the same set of quantum numbers $F$ and $p$, with increasing energy. The final state of the transition used to investigate the singlet contribution to the $b
^3\Sigma^+$ state wave function in section \ref{section:b-a} is highlighted in red.}

\begin{tabular}{cccccc|cccccc}
\hline\hline

$E$ (cm$^{-1}$)& $g_F$& $N$& $F$& $p$& $n$ & $E$ (cm$^{-1}$)& $g_F$& $N$& $F$& $p$& $n$\\ \hline
 44804.4031&$-0.661$&0&2&$+1$&1&44807.7613&$-0.298$&2&4&+1&2\\
 44804.4305&$-1.000$&0&1&$+1$&1&44807.7630&$-0.517$&2&1&+1&2\\
 44804.5235&$~~0.327$&0&2&$+1$&2&44807.7685&$~~0.000$&2&0&+1&1\\
 44804.5268&$~~0.170$&0&3&$+1$&1&44807.7838&$-0.228$&2&2&+1&4\\
 44804.6622&$~~0.663$&0&3&$+1$&2&44807.7868&$-0.291$&2&3&+1&4\\
 44804.6928&$~~0.500$&0&4&$+1$&1&44807.7883&$~~0.261$&2&1&+1&3\\
 44805.5139&$-0.617$&1&2&$-1$&1&44807.8599&$~~0.000$&2&0&+1&2\\
 44805.5217&$-0.411$&1&3&$-1$&1&44807.8657&$~~0.517$&2&1&+1&4\\
 44805.5282&$-0.991$&1&1&$-1$&1&44807.8704&$~~0.551$&2&1&+1&5\\
 44805.5454&$-0.712$&1&1&$-1$&2&44807.8731&$~~0.140$&2&4&+1&3\\
 44805.5479&$-0.461$&1&2&$-1$&2&44807.8749&$~~0.090$&2&5&+1&1\\
 44805.5540&$~~0.000$&1&0&$-1$&1&44807.8792&$~~0.308$&2&2&+1&5\\
 44805.6297&$~~0.703$&1&1&$-1$&3&44807.8844&$~~0.323$&2&2&+1&6\\
 44805.6344&$~~0.292$&1&2&$-1$&3&44807.8917&$~~0.179$&2&3&+1&5\\
 44805.6373&$~~0.196$&1&3&$-1$&2&44807.8930&$~~0.153$&2&3&+1&6\\
 44805.6396&$~~0.117$&1&4&$-1$&1&\color{red}44807.8966&\color{red}$~~0.088$&\color{red}2&\color{red}4&\color{red}+1&\color{red}4\\
 44805.6548&$~~0.341$&1&2&$-1$&4&44808.0120&$~~0.454$&2&4&+1&5\\
 44805.6615&$~~0.172$&1&3&$-1$&3&44808.0151&$~~0.470$&2&3&+1&7\\
 44805.7708&$~~0.632$&1&3&$-1$&4&44808.0237&$~~0.397$&2&5&+1&2\\
 44805.7845&$~~0.497$&1&4&$-1$&2&44808.0253&$~~0.569$&2&2&+1&7\\
 44805.7958&$~~0.779$&1&2&$-1$&5&44808.0345&$~~1.188$&2&1&+1&6\\
 44805.8053&$~~0.486$&1&4&$-1$&3&44808.0457&$~~0.380$&2&5&+1&3\\
 44805.8142&$~~0.400$&1&5&$-1$&1&44808.0468&$~~0.415$&2&4&+1&6\\
 44805.8216&$~~0.578$&1&3&$-1$&5&44808.0529&$~~0.333$&2&6&+1&1\\
 44807.7541&$-0.342$&2&3&$+1$&3&44808.0530&$~~0.499$&2&3&+1&8\\
 44807.7565&$-0.416$&2&2&$+1$&3&44808.0597&$~~0.777$&2&2&+1&8\\

\hline

\end{tabular}

\end{table}


\section{\label{section:b-a}The $b^3\Sigma^+, v'=0\leftarrow a^3\Pi, v''=0$ Transition}

The $b-a$ bands have a diagonal Franck-Condon matrix and the $b^3\Sigma^+, v'=0\leftarrow a^3\Pi, v''=0$ band has a Franck-Condon factor of 0.994. Its natural linewidth is 100 times smaller than the natural linewidth of the strong $A^1\Pi, v'=0\leftarrow X^1\Sigma^+, v''=0$ transition near 227.5~nm. Laser cooling AlF molecules on the $b-a$ transition could therefore reach temperatures far below the Doppler limit of the strong $A-X$ transition. The vibrational branching to $a^3\Pi, v=1$ is small and, if addressed with a repump laser, a molecule could scatter about 1000 photons before being pumped into $a^3\Pi, v=2$. However, the radiative decay from the $b$ state to multiple $J$ levels in all three spin-orbit manifolds of the $a$ state is allowed. This results in a large number of rotational branches that must be addressed to close the optical cycle \cite{Nolan1936}. In addition, the hyperfine structure in $a$ is large compared to the linewidth of the transition.  Both effects make laser-cooling of AlF, using an optical transition in the triplet manifold very challenging. This is in stark contrast to the strong $A-X$ transition, for which all Q-lines are rotationally closed and for which all hyperfine levels of a given rotational level in the $X$ state lie within the natural linewidth. 

Here, we demonstrate that laser-induced fluorescence spectroscopy of the $b^3\Sigma^+, v'=0\leftarrow  a^3\Pi, v''=0$ transition can be used to efficiently detect $a^3\Pi$ molecules with hyperfine resolution. In this section, we show that this method works well to detect molecules in all three $\Omega$ manifolds to improve two important spectroscopic constants of the $a^3\Pi$ state: the spin-orbit ($A$) and spin-spin ($\lambda$) interaction parameter, which determine the relative spacing of the three $\Omega$ manifolds in $a^3\Pi$. The effective fine structure Hamiltonian is

\begin{equation}
H=AL_zS_z+\frac{2}{3}\lambda(3S^2_z-\mathbf{S}^2).
\end{equation}
The interval between the $\Omega=0$ and $\Omega=1$ manifolds is approximately $A-2\lambda$, while the $\Omega=1$ and $\Omega=2$ manifolds are about $A+2\lambda$ apart. 

For this experiment we use the supersonic molecular beam setup introduced in section \ref{section:overview}, with an additional LIF detector installed between the excitation region and the TOF-MS. After passing through the skimmer, the molecules are excited on a specific rotational line of the $a^3\Pi, v'=0\leftarrow X^1\Sigma^+, v''=0$ band to one of the three spin-orbit manifolds, using a frequency-doubled pulsed dye amplifier (PDA), which is seeded by a cw titanium sapphire laser. About 30~cm further downstream, a cw ring dye laser intersects the molecular beam orthogonally and is scanned over a rotational line of the $b^3\Sigma^+, v'=0 \leftarrow a^3\Pi, v''=0$ band. The laser frequency is stabilised and scanned with respect to a frequency-stabilised and calibrated HeNe reference laser (SIOS SL 03), using a scanning transfer cavity. The wavelength is recorded with an absolute accuracy of 10~MHz using the wavemeter, calibrated by the same HeNe laser.

\begin{figure}[!htb]
\centering
\includegraphics[width=\linewidth]{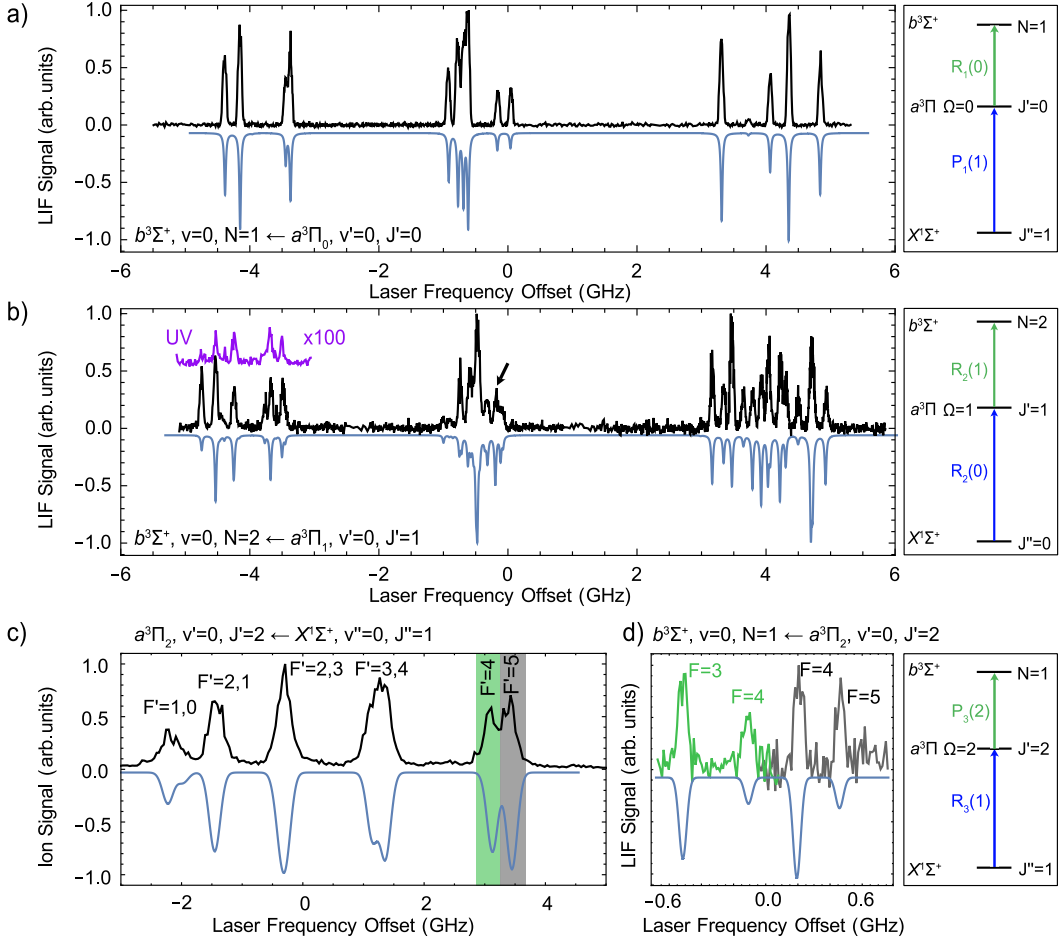}
\caption{\label{fig:Figure_6} LIF excitation spectra of the $b-a$ transition. Panels a) and b) show spectra of the $b^3\Sigma^+, v'=0, N'=1 \leftarrow a^3\Pi_0, v''=0, J''=0$ and $b^3\Sigma^+, v'=0, N'=2 \leftarrow a^3\Pi_1, v''=0, J''=1$  transitions, respectively. The spectrum in the inset labelled UV is detected by recording the emission on the $b^3\Sigma^+, v=0, N=2 \rightarrow X^1\Sigma^+, v''$ bands as a function of the excitation frequency.  c) Hyperfine resolved spectrum of the R$_3$(1) line of the $a^3\Pi, v'=0\leftarrow X^1\Sigma^+, v''=0$ band, showing the large hyperfine splitting in $a^3\Pi_2$. d) Part of the hyperfine resolved spectrum of the $b^3\Sigma^+, v'=0, N'=1 \leftarrow a^3\Pi_2, v''=0, J''=2$ transition. The low (high) frequency part of this spectrum was recorded after excitation to the $F=4$ ($F=5$) component of the R$_3$(1) line of the $a^3\Pi\leftarrow X^1\Sigma^+$ transition. a) - d): The relevant energy level scheme is displayed next to each spectrum. Since $J$ is not well-defined in the $b$ state, the $b^3\Sigma^+ \leftarrow a^3\Pi$ transitions are labelled with $\Delta(NJ)_{\mathcal{F'}}(J')$. The measured spectra are printed in black, while the simulated spectra are shown in blue and inverted.}
\end{figure}

Figure \ref{fig:Figure_6} shows hyperfine-resolved LIF excitation spectra of the $b^3\Sigma^+, v'=0\leftarrow a^3\Pi, v''=0$ band. Figure \ref{fig:Figure_6}a shows the hyperfine spectrum of the $b^3\Sigma^+, v'=0, N'=1 \leftarrow a^3\Pi_0, v''=0, J''=0$ lines. The positive parity $\Lambda$-doublet level of $a^3\Pi_0, J=0$ has only two hyperfine components with total angular momentum quantum numbers $F=2$ and $F=3$ that are only split by about 3~MHz, much smaller than the residual Doppler broadening in the molecular beam. Therefore, this $b-a$ spectrum directly reflects the energy level structure in the $b$ state and is identical to the one of the $b-X$ spectrum shown in Figure \ref{fig:Figure_5}b. 

The spectrum of the $b^3\Sigma^+, v'=0, N'=2 \leftarrow a^3\Pi_1, v''=0, J''=1$ transition, presented in Figure \ref{fig:Figure_6}b, shows a richer structure, and no longer directly reflects the energy level structure in the $b$ state. The total span of the hyperfine splitting in the $a^3\Pi_1, v=0, J=1$ level is about 500 MHz \cite{Truppe2019}, and contributes to the complexity of this spectrum.

Optical pumping to the $a^3\Pi_2, v=0$ state is challenging because the $R_3(1)$ transition of the $a^3\Pi_2, v'=0 \leftarrow X^1\Sigma^+, v''=0$ band is about 1000 times weaker than the corresponding $R_2(1)$ line to the $a^3\Pi_1$ state. Only a small fraction of the ground-state molecules produced in the source is transferred to the $a^3\Pi_2, v=0$ level, even with a PDA pulse energy of about 10 mJ in beam with a $e^{-2}$ waist radius of 1.5 mm. Figure \ref{fig:Figure_6}c shows the R$_3(1)$ line of the $a^3\Pi_2, v'=0 \leftarrow X^1\Sigma^+, v''=0$ band, recorded by scanning the seed laser of the frequency-doubled PDA, followed by (1+2)-REMPI and TOF-MS detection. The hyperfine structure in $a^3\Pi_2, v=0$ is very large and it is possible to resolve six of the ten hyperfine levels using a narrow-band pulsed laser. This enables us to populate a specific hyperfine component of the $a^3\Pi_2, v=0, J=2$ level, which is then followed by cw excitation to the $b$ state and LIF detection. Figure \ref{fig:Figure_6}d shows two LIF spectra that originate from two different hyperfine states in $a^3\Pi_2$, indicated by the two colours. The green spectrum originates from the $a^3\Pi_2, v=0, J=2, F=4$ level and the grey spectrum from the $a^3\Pi_2, v=0, J=2, F=5$ level. The line-centres are determined by fitting Gaussians to the spectral lines.

The hyperfine structure and $\Lambda$-doubling of the $a^3\Pi$ state has been investigated extensively in our previous study and is known to kHz precision \cite{Truppe2019}. However, the constants $A$ and $\lambda$ of the $a^3\Pi$ state could only be determined with an accuracy of about 200 MHz, due to the finite bandwidth of the pulsed laser and the lower accuracy of the wavemeter that was used in that study to measure the relative energy of the $\Omega$ manifolds. Here, we take advantage of the reduced linewidth in the cw LIF spectra of the $b-a$ transition, in combination with the increased accuracy of our new wavemeter, to improve this measurement and therefore the spectroscopic constants $A$ and $\lambda$. By using the hyperfine constants for the $a^3\Pi$ state from reference \cite{Truppe2019} in combination with the parameters for the $b$ state, listed in Table \ref{table:Table_1}, the $b-a$ spectra can be simulated, and new values for $A$ and $\lambda$ are derived. The simulated spectra, using the full Hamiltonian, including the hyperfine interactions in both states are shown in Figure \ref{fig:Figure_6} as blue, inverted curves. The uncertainty of the improved spectroscopic constants, listed in Table \ref{table:Table_3}, is reduced by nearly two orders of magnitude compared to the ones presented previously \cite{Truppe2019}. $E_0$ is the term energy of the $a$ state in the absence of rotation, fine and hyperfine structure, calculated by using the term energy of the $b$ state determined in the previous section. This means that the gravity centre, i.e, the position of the $a^3\Pi_1, v=0, J=1$ level in the absence of hyperfine structure, relative to the $X^1\Sigma^+, J=0$ level is at 27255.1737(5)~cm$^{-1}$. 

It is also possible to record the LIF excitation spectra by detecting the weak UV fluorescence, on the $b^3\Sigma^+, v'=0 \rightarrow X^1\Sigma^+, v''$ bands which occurs predominantly at 223.1~nm. Part of the spectrum displayed in \ref{fig:Figure_6}b has also been recorded this way and is shown by the inset labelled `UV'. To measure the ratio of the emission occurring in the UV, relative to the emission occurring in the visible, we lock the excitation laser to the resonance indicated by the arrow in \ref{fig:Figure_6}b and average the signal over 1000 shots. To distinguish the two wavelengths we use two different PMTs: a UV sensitive one with a specified quantum efficiency of $0.35\pm0.05$ at 223.1~nm and negligible sensitivity for wavelengths $>350$~nm and a second PMT with a specified quantum efficiency of $0.1\pm0.015$ at 569~nm that is sensitive in the range of $200-800$~nm. The quantum efficiencies of the PMTs are taken from the data sheet and are not calibrated further. However, three identical UV PMTs give the same photon count rate to within 15\% which we take as the systematic uncertainty. Both PMTs are operated in photon-counting mode, counting the number of UV and visible photons, $n_{\textrm{uv}}$ and $n_{\textrm{vis}}$, respectively. The visible PMT is combined with a band-pass interference filter to block the UV fluorescence and a small amount of phosphorescence on the $a^3\Pi, v'=0 \rightarrow X^1\Sigma^+, v''=0$ transition near 367~nm. The measured transmission of the filter at 569~nm is 0.57. We combine the slightly different transmission through the imaging optics, the different detector efficiencies and the filter transmission into the total detection efficiencies $\eta_{\textrm{uv}}$ and $\eta_{\textrm{vis}}$. Including these values, we measure a ratio of 
\begin{equation}
    R_b=\frac{n_{\textrm{uv}}}{n_{\textrm{vis}}} \frac{\eta_{\textrm{vis}}}{\eta_{\textrm{uv}}}=(4.3\pm 1.3)\times 10^{-3}.
\end{equation} 
This is the result of multiple experimental runs with a statistical error that is significantly smaller than the uncertainty in the quantum efficiencies of the PMTs. An additional 20\% systematic uncertainty is added to the total error budget because we do not correct for the spherical and chromatic aberration of the fluorescence detector. However, simulations using ray-tracing software show a typical difference in imaging efficiency for the two wavelengths of about 10\%. The value for $R_b$ is identical to the ratio of the Einstein $A$-coefficients for $b \rightarrow X$ and $b \rightarrow a$ emission, i.e. $R_b$ = $A_{b,X}/A_{b,a}$. The specific level in the $b^3\Sigma^+, v=0$ state for which this ratio of Einstein $A$-coefficients is determined, is the positive parity level with $N=2$, $F=4$, $n=4$ which is almost a pure $f$-level, with 42.3\% $\mathcal{F}_3$ and 57.4\% $\mathcal{F}_1$ character, and only 0.3\% $\mathcal{F}_2$ character. The level is highlighted in red in table \ref{table:Table_2} and figure \ref{fig:Figure_4}.

\begin{table}[ht]
\centering
\caption{\label{table:Table_3} Spectroscopic constants for the $a^3\Pi, v=0$ state of AlF, determined from a fit to the spectra. $E_0$ is the energy of the electronic state in the absence of rotation, fine- and hyperfine structure relative to the $X^1\Sigma^+, v=0, J=0$ level (in cm$^{-1}$). The spin-orbit coupling constant, $A$, and the spin-spin interaction constant, $\lambda$, are given in MHz. Their respective values from our earlier publication \cite{Truppe2019} are also reproduced for comparison.}
\begin{tabular}{cllll} \hline\hline
 Parameter & Value & SD & Value \cite{Truppe2019} & SD\\
 \hline
 $E_0$          &   27253.0507      &   0.0005  & 27253.04&0.01\\
 $A$            &   1421089         &   5       &1420870&210\\
 $\lambda$      &   2766            &   4       &2659&33\\
 \hline\hline
\end{tabular}
\end{table}

\section{The $A^1\Pi\rightarrow a^3\Pi$ Bands}
To probe the amount of triplet wave function that is mixed into the $A^1\Pi$ state directly, we measure the ratio $R_A$ of the number of fluorescence photons emitted on the $A^1\Pi\rightarrow a^3\Pi$ and on the $A^1\Pi\rightarrow X^1\Sigma^+$ transition. The value for $R_A$ is identical to the ratio of the Einstein $A$-coefficients for $A \rightarrow a$ and $A \rightarrow X$ emission, i.e., $R_A$ = $A_{A,a}/A_{A,X}$. The value for $R_A$ gives the loss from the main laser cooling cycle due to electronic branching to the $a^3\Pi$ state. Previously, we measured this electronic branching ratio on the $A\rightarrow a$ bands indirectly to be at the $10^{-7}$ level, by comparing the absorption-strength of the $A^1\Pi, v=0 \leftarrow a^3\Pi_0, v'=0$ transition relative to the absorption-strength of the $A^1\Pi, v=0 \leftarrow X^1\Sigma^+, v''=0$ transition \cite{Truppe2019}. In that analysis we only took into account the amount of singlet character in the wave function of the $a$ state to deduce the strength of the $A \leftarrow a$ transition. We did not account for the (then unknown) amount of triplet character in the wave function of the $A$ state due to the interaction of the $A$ and $b$ states, which, as we will see here, turns out to be the dominant effect.

\begin{figure}[!htb]
\centering
\includegraphics{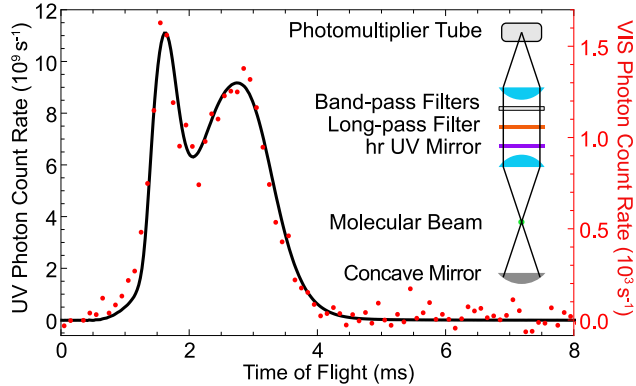}
\caption{\label{fig:Figure_7} The LIF emitted in the UV (227.5~nm, black) and VIS (599~nm, red) as a function of time when the UV laser frequency is locked to the Q(1) line of the $A^1\Pi, v'=0 \leftarrow X^1\Sigma^+, v''=0$ transition. The inset shows the configuration of the fluorescence detector used for the VIS experiment.}
\end{figure}

To measure such a small branching ratio directly, we use the buffer gas molecular beam source in combination with optical cycling on the Q(1) line of the $A-X$ transition. The small amount of visible fluorescence is isolated from the strong UV fluorescence by a high-reflectivity UV mirror, which is transparent in the visible, in combination with a long-pass and two bandpass filters in front of the PMT. The transmission band of the filters is chosen such that only wavelengths that cover the $A
^1\Pi, v'=0\rightarrow a^3\Pi, v''=0$ transition are detected by the PMT. The mirror reduces the UV fluorescence that is incident on the spectral filters to the $10^{-6}$ level, suppressing their broadband phosphorescence when irradiated with UV light. The transmission of each optical element is measured individually at 227.5~nm and 599~nm, using laser light and a calibrated photo diode. The total detection efficiency, accounting for the transmittances and PMT responses becomes $\eta_{uv}=0.2 \pm 0.05$ for UV photons and $\eta_{vis}= 0.044 \pm 0.01$ for visible photons in the range of $596-604$~nm. The ratio of Einstein coefficients becomes
\begin{equation}
\label{eq:ra}
    R_A=\frac{n_{\textrm{vis}}}{n_{\textrm{uv}}}\frac{\eta_{\textrm{uv}}}{\eta_{\textrm{vis}}}=(6\pm2)\times10^{-7},
\end{equation}
where $n_{\textrm{vis}}$ and $n_{\textrm{uv}}$ are the number of photons detected in the visible and UV, respectively. A typical measurement is presented in Figure \ref{fig:Figure_7}. The molecules are optically pumped on the Q(1) line of the $0-0$ band of the $A-X$ transition and the LIF is imaged and detected by two different PMTs. The majority of the fluorescence is emitted in the UV and imaged onto the UV sensitive PMT. The PMT is operated in current mode, which is converted into a voltage, amplified and read into the computer. We calibrate this PMT output voltage against the output of the PMT in photon-counting mode for low incident light intensities. The small fraction of the LIF that is emitted in the visible is shown as red dots. The two time of flight profiles are very similar, with the detected signal in the visible being $(1.3\pm 0.05)\times10^{-7}$ of the emission in the UV. By accounting for the different detection efficiencies for the two wavelengths the measured ratio is as given in equation \ref{eq:ra}. The total uncertainty in this measurement is dominated by the systematic uncertainty in the quantum efficiency of the two PMTs and by the uncertainty in the imaging efficiency for the two wavelengths as described in section \ref{section:b-a}. 

The measurement of $R_A$ is significantly more challenging than the measurement of $R_b$. This is because $n_{\textrm{uv}}$ is $10^7$ times larger than $n_{\textrm{vis}}$ and the phosphorescence of the optical elements typically occurs red-shifted, i.e., in the wavelength range of the weak fluorescence in the visible. This is in contrast to the measurement of $R_b$, for which $n_{\textrm{uv}}$ is about 200 times smaller than $n_{\textrm{vis}}$, and for which any background caused by phosphorescence of the optical elements is absent.

We also measure the ratio of the visible to the UV fluorescence subsequent to excitation on the Q(1) line of the $1-1$ band of the $A-X$ transition. Since the $A^1\Pi, v=1$ level is energetically much closer to the $b^3\Sigma^+, v=0$ level, one might expect a significantly larger fraction of visible fluorescence. However, we find that the two ratios $R_A$ are equal to within the 15\% uncertainty of the measurement.

\section{\label{section:perturbations}Spin-orbit Interaction Between the $A^1\Pi$ and $b^3\Sigma^+$ States}

\begin{figure}[!htb]
\centering
\includegraphics{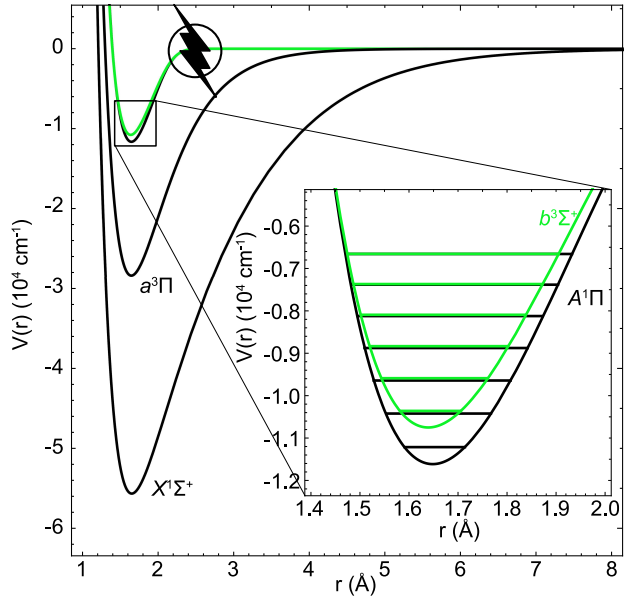}
\caption{\label{fig:Figure_8}Potential energy curves for the lowest singlet and triplet electronic states of AlF, using precise Expanded Morse Oscillator (EMO) functions. We obtain these EMO potentials by fitting to the point-wise RKR potentials generated by LeRoy's program \cite{LeRoy2017} and adjust the parameters to predict the vibrational levels with a high accuracy of 0.05~cm$^{-1}$ (using the non-perturbed or de-perturbed values from the appendix from \cite{Barrow1974}) and from \cite{Yousefi2018}. These potentials are much more precise than the simple Morse model we use in section \ref{section:perturbations}, but do not allow us to treat the two electronic states independently to extract the spin-orbit interaction. {\textit{Ab initio}} calculations indicate that the $A^1\Pi$ state has a barrier in the region marked with a flash, which cannot be reproduced with our EMO potentials. The inset shows a more detailed view of the $A^1\Pi$ and $b^3\Sigma^+$ potentials, with the vibrational levels indicated.}
\end{figure}

The observation of the $b^3\Sigma^+, v'=0\leftrightarrow X^1\Sigma^+, v''$ and of the $A^1\Pi, v'=0 \rightarrow a^3\Pi, v''$ intersystem bands reported here indicates that the wave function of the $b$ state of AlF contains a fraction of singlet character and that the $A$ state contains a fraction of triplet character. This is due to the spin-orbit interaction of the $b$ state with the nearby $A^1\Pi$ state. Figure \ref{fig:Figure_8} shows the potential energy curves for the lowest singlet and triplet electronic states of AlF. The inset shows a more detailed view of the $A^1\Pi$ and $b^3\Sigma^+$ states with their vibrational levels indicated. For low vibrational quantum numbers, the vibrational level $v_b$ in the $b$ state lies slightly above the vibrational level $v_A=v_b + 1$ in the $A$ state. The energy difference between $v_b$ and $v_A=v_b + 1$ decreases with increasing vibrational quantum numbers until the levels have just crossed and are nearly degenerate at $v_A=6$ ($v_b=5$), leading to a large perturbation of the rotational energy levels. This interaction has already been analysed by Barrow {\it et al.} in 1974, who introduced the parameter $H_0(v_A)$ to describe the effect of the spin-orbit interaction of specific pairs of vibrational levels, i.e., of $v_A=4, 5$ and $6$ with $v_b=3, 4$ and $5$, respectively \cite{Barrow1974}. They used the observed perturbation to determine the energy of the triplet manifold relative to the singlet manifold with an accuracy of 0.05~cm$^{-1}$. 

In this section we give a more general description of the spin-orbit interaction between the $A^1\Pi, v_A$ state and the $b^3\Sigma^+, v_b$ state. We deduce the expressions for $R_b$ and $R_A$ in terms of the spin-orbit coupling constant $A_{so}$ between the $A$ and the $b$ state, to experimentally determine the value for $A_{so}$. This is then compared to the value of $A_{so}$ that can be deduced from the $H_0(v_A)$ parameters as given by Barrow and co-workers \cite{Barrow1974}. The effect of the interaction on the rotational energy levels in the $A^1\Pi, v=0$ and $b^3\Sigma^+, v=0$ states is also discussed.

For the spin-orbit interaction of a $^1\Pi$ state with a $^3\Sigma^+$ state, the interaction terms between the $e$- and $f$-levels belonging to a given $J$ are given by \cite{1969Kovacsp}:

\begin{eqnarray}
H(^1\Pi, f | ^3\Sigma^+, \mathcal{F}_1) &=& \sqrt{\frac{J+1}{2J+1}} \frac{\xi}{\sqrt{2}} \\
H(^1\Pi, e | ^3\Sigma^+, \mathcal{F}_2) &=& - \frac{\xi}{\sqrt{2}} \\
H(^1\Pi, f | ^3\Sigma^+, \mathcal{F}_3) &=& \sqrt{\frac{J}{2J+1}} \frac{\xi}{\sqrt{2}}.
\end{eqnarray}
In the Born-Oppenheimer approximation, the parameter $\xi$ only depends on the vibrational wave functions of the coupled states and can be written in terms of the spin-orbit operator $H_{so}$ as
\begin{equation}
\label{eq:sointeraction}
\xi(v_A,v_b) = \bra{\Psi_{A,v_A}}H_{so}\ket{\Psi_{b,v_b}}=A_{so}\int\,\phi^{*}_{b,v_b}(\rho)\,\phi_{A,v_A}(\rho)\,d\rho = A_{so}\sqrt{q_{v_Av_b}}.
\end{equation}
The expressions $\phi_{b,v_b}(\rho)$ and $\phi_{A,v_A}(\rho)$ are the vibrational wave functions for the $b$ and $A$ state, respectively, and $\rho$ is the inter-nuclear distance between the Al and F atoms. The square of the expression for the integral is the Franck-Condon factor $q_{v_Av_b}$ between the $A, v_A$ and $b,v_b$ levels. The Franck-Condon matrix between the $A$ and $b$ states is very diagonal. The value of $q_{00}$ is very close to one and even though the $v=0$ levels of both states are about $(E_{b,0} - E_{A,0})$ = 855~cm$^{-1}$ apart, the interaction between these levels dominates. The value of $q_{10}$ is much smaller than one, but as the $v_b=0$ and $v_A=1$ levels are only $(E_{b,0} - E_{A,1})$ = 63~cm$^{-1}$ apart, this interaction also has to be taken into account. The interaction with all the other vibrational levels can be neglected.

\subsection{Intensities of the Intersystem Bands}
The spin-orbit interaction between the $A$ state and the $b$ state mixes their wave functions and causes a shift of the rotational levels. If the interacting levels are much further apart than the magnitude of the interaction terms, we can calculate the effect of the interaction using first order perturbation theory. In this case, the contribution to the wave function of the $A^1\Pi, v=0$ ($b^3\Sigma^+, v=0$) state due to the interaction with the $b$ ($A$) state is given by the interaction term divided by the energy separation of the interacting levels. The wave function of the $A$ state can be written as:
\begin{equation}
\ket{\Psi'_{A,0}}=C^s_{A,0}\ket{\Psi_{A,0}}+C^t_{A,0;b,0} \ket{\Psi_{b,0}},
\end{equation}
where $|C^s_{A,0}|\approx 1$, is the total amount of singlet character and where $|C^t_{A,0;b,0}|$ $\equiv$ $|C^t_{A,0}|$ is the total amount of triplet character in the wave function of the $A$ state. It is readily seen from the interaction terms given above that the triplet contribution to the wave functions of the $e$- and $f$-levels is equal, does not depend on $J$, and is given by
\begin{equation}
|C^t_{A,0}| = \frac{A_{so}\sqrt{q_{00}}}{\sqrt{2}(E_{b,0} - E_{A,0})}.
\end{equation}
The wave function of the $b$ state can be written as:
\begin{equation}
\ket{\Psi'_{b,0}}=C^t_{b,0}\ket{\Psi_{b,0}} + C^s_{b,0;A,0}(J,\mathcal{F}_i) \ket{\Psi_{A,0}} +
C^s_{b,0;A,1}(J,\mathcal{F}_i) \ket{\Psi_{A,1}},
\end{equation}
where $|C^t_{b,0}|\approx 1$ is the total amount of triplet character and where $\sqrt{|C^s_{b,0;A,0}(J,\mathcal{F}_i)|^2 + |C^s_{b,0;A,1}(J,\mathcal{F}_i)|^2}$ $\equiv$ $|C^s_{b,0}(J,\mathcal{F}_i)|$ is the total amount of singlet character in the wave function of the $b$ state. The singlet contribution to the wave function of the $b$ state depends on the $(J, \mathcal{F}_i)$ ($i=1,2,3$) level. For the positive parity, $N=2$, $F=4$, $n=4$ level that we used for the measurement of $R_b$ we find, using the weights $W^J$ as indicated in figure \ref{fig:Figure_4} ($W^J_1=42.3\%$, $W^J_2=0.3\%$ and $W^J_3=57.4\%$),
\begin{equation}
\label{eq:cb}
|C^s_{b,0}| = 0.69\frac{A_{so}}{\sqrt{2}} \sqrt{\frac{q_{00}}{(E_{b,0} - E_{A,0})^2} + \frac{q_{10}}{(E_{b,0} - E_{A,1})^2}}.
\end{equation}

The expression for $R_A$ can now be rewritten as:
\begin{equation}
R_A = \frac{A_{A,a}}{A_{A,X}} = \frac{\abs{\bra{\Psi\smash{'}_{A,0}}\mu(r)\ket{\Psi_{a,0}}}^2 \lambda_{A,X}^3}{\abs{\bra{\Psi\smash{'}_{A,0}}\mu(r)\ket{\Psi_{X,0}}}^2 \lambda_{A,a}^3} = |C^t_{A,0}|^2 \frac{A_{b,a} \lambda_{b,a}^3 }{A_{A,X} \lambda_{A,a}^3} = \frac{|C^t_{A,0}|^2}{116}, 
\end{equation}

where $\mu(r)$ is the electronic transition dipole moment, $\lambda_{b,a}=569$~nm, $\lambda_{A,a}=599$~nm are the wavelengths of the $b-a$ and $A-a$ transitions, respectively. The ratio of the Einstein $A$-coefficients is calculated from the experimentally known lifetimes of the $A$ state (1.90~ns, \cite{Truppe2019}) and of the $b$ state (190~ns, Section \ref{section:overview}). The expression for $R_b$ can now be rewritten as:
\begin{equation}
R_b = \frac{A_{b,X}}{A_{b,a}} = \frac{{\sum_{i=0}^{1}\abs{\bra{\Psi\smash{'}_{b,0}}\mu(r)\ket{\Psi_{X,i}}}}^2 \lambda_{b,a}^3}{\abs{\bra{\Psi\smash{'}_{b,0}}\mu(r)\ket{\Psi_{a,0}}}^2 \lambda_{b,X}^3} = |C^s_{b,0}|^2 \frac{A_{A,X} \lambda_{A,X}^3}{A_{b,a} \lambda_{b,X}^3} = 
106 |C^s_{b,0}|^2,
\end{equation}
where $\lambda_{A,X}$ = 227.5~nm and $\lambda_{b,X}$ = 223~nm. The sum in the numerator results in the two terms given in equation \ref{eq:cb}, because the Franck-Condon matrix between the $A$ and $X$ states is highly diagonal. \\

The parameter $H_0(v_A)$, introduced by Barrow and co-workers, describes the effect of the spin-orbit interaction between two specific vibrational levels and is equivalent to 
\begin{equation}
    H_0(v_A) = \frac{\xi(v_A,v_b=v_A-1)}{\sqrt{2}} = A_{so} \sqrt{\frac{q_{v_A v_A-1}}{2}}.
\end{equation}
Barrow and co-workers did not discuss the dependence of the values for $H_0(v_A)$ on the square root of the Franck-Condon factors, and they therefore did not extract a single value for the spin-orbit interaction parameter $A_{so}$ from the three values of $H_0(v_A)$ that they reported \cite{Barrow1974}. 

In the following, we determine the Franck-Condon factors $q_{v_Av_b}$ and $A_{so}$ from the measured $H_0(v_A)$ values \cite{Barrow1974}, and then use these Franck-Condon factors to extract a value for $A_{so}$ from our measurements of $R_A$ and $R_b$. For this, a Morse potential is fitted to the term-values of the vibrational levels in the $A$ and $b$ state, as listed in \cite{Barrow1974}. The parameters of the Morse potential are optimised to reproduce the measured vibrational levels to better than 0.3~cm$^{-1}$. This optimization is independent of the equilibrium distance $r_e$. Next, the difference between the equilibrium distances of the $A$ and $b$ state, $\Delta r_e=r_e(A) - r_e(b)$, is optimised such that the vibrational level dependence of $\sqrt{q_{v_Av_A-1}}$ agrees with the observed vibrational level dependence of $H_0(v_A)$. We find the best agreement for $\Delta r_e= 0.0046\textrm{\AA}$, about half the value for $\Delta r_e$ extracted from the reported values for $B_e$ in the $A$ and $b$ state \cite{Barrow1974} of 0.0094~$\textrm{\AA}$. Finally, we fit to the experimentally determined rotational constants reported in \cite{Barrow1974} using only $r_e(A)$ as a fit parameter. The data is reproduced to within 1\% for $r_e(A)$ = 1.63098~$\textrm{\AA}$. Considering the simple model for the potentials, this is an excellent agreement. The value for $A_{so}$ determined from this fitting procedure is $A_{so}= 8.9$~cm$^{-1}$ and the corresponding Franck-Condon matrix is shown in Table \ref{table:Table_4}. The uncertainty in the value of $A_{so}$ is difficult to determine, as the main contribution to the total uncertainty stems from the assumption that the potentials can be approximated by Morse potentials. We estimate this uncertainty to be at least 2.5~cm$^{-1}$.

\begin{table}[ht]
\centering
\caption{\label{table:Table_4} Calculated Franck-Condon factors between the $A^1\Pi$ state and the $b^3\Sigma^+$ state. The elements highlighted in red are used to predict the three measured values for $H_0(v_A)$ given in \cite{Barrow1974}}
 \begin{tabular}{lcccccccc} \hline
$v_A \backslash v_b$ &   0        &  1       & 2         &   3       &   4       &   5       &   6       \\
\hline
0           &   0.9936   &  0.0012  & 0.0051    &   0.0002  &   0.0000  &   0.0000  &   0.0000  \\
1           &   0.0010   &  0.9831  & 0.0005    &   0.0144  &   0.0010  &   0.0000  &   0.0000  \\
2           &   0.0051   &  0.0003  & 0.9650    &   0.0000  &   0.0264  &   0.0031  &   0.0000  \\
3           &   0.0003   &  0.0142  & 0.0002    &   0.9364  &   0.0024  &   0.0391  &   0.0073  \\
4           &   0.0001   &  0.0008  & 0.0267    &   \color{red}0.0023  &   0.8950  &   0.0102  &   0.0498  \\
5           &   0.0000   &  0.0004  & 0.0014    &   0.0425  &   \color{red}0.0078  &   0.8390  &   0.0260  \\
6           &   0.0000   &  0.0001  & 0.0009    &   0.0019  &   0.0615  &   \color{red}0.0168  &   0.7680  \\
\hline
 \end{tabular}
\end{table}

Using the information from the Franck-Condon matrix shown in Table \ref{table:Table_4}, the experimental value of $R_A = (6 \pm 2)\times 10^{-7}$ implies a value for $A_{so}= (10 \pm 2)$~cm$^{-1}$. The value of $R_b = (4.3 \pm 1.3) \times 10^{-3}$ implies a value for $A_{so} = (10.4 \pm 1.5)$~cm$^{-1}$. These two values for A$_{so}$, as well as the value for $A_{so}$ extracted from the $H_0(v_A)$ values, all overlap within their error bars, yielding a final experimental value for $A_{so} = (10.0 \pm 1.1)$~cm$^{-1}$.

As mentioned in the previous section, we observe that the ratio of the visible to the UV fluorescence that is emitted from the $A^1\Pi, v=1$ level is equal to that from the $A^1\Pi, v=0$ level. Based on the derivation presented in this section we expect that the fractional emission in the visible from the $A^1\Pi, v=1$ level is larger than the emission from the $A^1\Pi, v=0$ level by a factor of
\begin{equation}
    \left(\frac{E_{b,0}-E_{A,0}}{E_{b,1}-E_{A,1}}\right)^2\frac{q_{11}}{q_{00}}+\left(\frac{E_{b,0}-E_{A,0}}{E_{b,0}-E_{A,1}}\right)^2\frac{q_{10}}{q_{00}}=1.23,
\end{equation}
where the energy difference $(E_{b,1} - E_{A,1})=834$~cm$^{-1}$, and where the values for the Franck-Condon factors are taken from Table \ref{table:Table_4}. The predicted 23\% increase in visible fluorescence is consistent with the experimental observation, within the experimental uncertainty of 15\%. This highlights that the spin-orbit mixing is dominated by the interaction between vibrational levels that have the same vibrational quantum number.

\subsection{Effect on the Fine Structure in the $b^3\Sigma^+, v=0$ State}
In first order perturbation theory, the shift of the energy levels is given by the square of the interaction matrix elements, divided by the energy separation of the interacting levels. The interaction is repulsive, and for the $A^1\Pi, v=0$ state all rotational levels shift downwards by $A^2_{so}q_{00}/[2(E_{b,0} - E_{A,0})]$. Such an overall shift is difficult to determine, and is absorbed in the value for the term-energy. In the $b^3\Sigma^+, v=0$ state, the $e$-levels will be shifted upward by about the same amount, but the $f$-levels will have a lower, $J$-dependent shift.`Curiously', Hebb wrote originally in 1936, the shift of the levels in a $^3\Sigma^+$ state due to the spin-orbit interaction with a $^1\Pi$ state has the same $J$-dependence as the shift due to the spin-spin and spin-rotation interaction, and both effects cannot be distinguished \cite{Hebb1936}. The origin of the spin-spin and spin-rotation interaction in a $^3\Sigma$ state has first been described in a classic paper by Kramers \cite{Kramers1929}, and more general expressions have been given soon after that by Schlapp \cite{Schlapp1937}. Normally, the spin-spin and spin-rotation interactions are expected to be the dominant effects and the spin-orbit interaction with a nearby $^1\Pi$ state is expected to be only a second order correction. Both effects add up, and this means that the values for $\lambda$ and $\gamma$ as found from fitting the energy levels in the $b^3\Sigma^+, v=0$ state should actually be interpreted as
\begin{eqnarray}
\lambda &=& \lambda_{ss} + \frac{A^2_{so}q_{00}}{4(E_{b,0}-E_{A,0})}\\
\gamma &=& \gamma_{sr} + \frac{B A^2_{so}q_{00}}{2(E_{b,0}-E_{A,0})^2}
\end{eqnarray}
where $\lambda_{ss}$ and $\gamma_{sr}$ describe the contribution due to the spin-spin and spin-rotation interaction in the $b^3\Sigma^+$ state, respectively, and where the additional terms describe the contribution due to the spin-orbit interaction with the nearby $A^1\Pi$ state.\\

When we take the final experimental value for $A_{so}$, then the spin-orbit contribution to $\lambda$ amounts to about +900 MHz. We conclude therefore that the value for $\lambda_{ss}$ in the $b^3\Sigma^+, v=0$ state is about -1800~MHz, and that about half of this value is cancelled by the spin-orbit interaction with the nearby $A^1\Pi$ state.
The spin-orbit contribution to $\gamma$ amounts only to +1.1~MHz, and this contribution can be neglected. For higher vibrational levels in the $b^3\Sigma^+$ state a slightly different behaviour is expected. The spin-orbit contribution to $\lambda$ remains about +900~MHz for $v_b = 0-3$, but increases to about +1200~MHz for $v_b=4$ due to the near-resonant interaction with the $v_A=5$ level. For the $v_b=6$ level, the near-resonant contribution to $\lambda$ due to spin-orbit interaction with the $v_A=7$ level is negative, reducing the total spin-orbit contribution to $\lambda$ to about +600~MHz. The $v_b=5$ level is special, as the $\mathcal{F}_3$ levels of the $b$-state and the $f$-levels of $A, v_A=6$ cross, between $J=1$ and $J=2$, making an interpretation in terms of a contribution to $\lambda$ less meaningful. It is interesting to note that this crossing causes the $J$-levels that belong to low $N$-values in the $b$-state to be split considerably further than for the $b^3\Sigma^+, v=0$ state, making $J$ a good quantum number. This strong interaction opens a `doorway' to efficiently drive transitions between the singlet and triplet manifolds \cite{Blokland2011,Bartels2012} and when $J$ is a good quantum number, this can be done highly rotationally selective. 

\section{Conclusion}
We investigated the $b^3\Sigma^+$ state of AlF and the spin-orbit interaction between this lowest electronically excited state in the triplet system and the first electronically excited singlet state, the $A^1\Pi$ state. First, we presented a low-resolution rotational spectrum of the $b^3\Sigma^+, v'=0 \leftarrow a^3\Pi_1, v''=0$ transition and determined the radiative lifetime of the $b^3\Sigma^+, v=0$ state to be 190(2)~ns. Molecules in the $a^3\Pi, v=0$ state can be efficiently detected using a (1+2)-REMPI scheme via the $b^3\Sigma^+, v=0$ state even at relatively low laser intensity. Then, we used cw laser induced fluorescence excitation spectroscopy of the $b^3\Sigma^+, v'=0 \leftarrow X^1\Sigma^+, v''=1$ transition to determine the fine and hyperfine structure of the $b^3\Sigma^+, v=0$ state with a precision of about 10~MHz. The eigenvalues of the hyperfine Hamiltonian have been fitted to the experimentally determined line positions and all relevant spectroscopic constants have been determined. Hyperfine-resolved LIF spectra of the $b^3\Sigma^+, v'=0 \leftarrow a^3\Pi, v''=0$ band, originating from all three spin-orbit manifolds in $a^3\Pi$, were used to improve the spin-orbit ($A$) and spin-spin ($\lambda$) interaction parameters that determine the relative spacing of the three $\Omega$ manifolds in the $a$ state. 
Despite the highly-diagonal Franck-Condon matrix of the $b-a$ transition, laser cooling of AlF in the triplet manifold is challenging, due to the large number of rotational branches. This is further complicated by the spin-orbit interaction between the $A^1\Pi$ state and the $b^3\Sigma^+$ state, which is concluded to be governed by an interaction parameter $A_{so}$ of about 10~cm$^{-1}$. The spin-orbit interaction mixes up to about 1\% of the wave function of the $A^1\Pi$ state into the wave function of the hyperfine levels in the $v=0$ level of the triplet state; the exact amount of mixing depends on the $J,\mathcal{F}_i$ ($i=1,2,3$) character of the hyperfine levels in the $b^3\Sigma^+$ state. By the same mechanism, about 1\% of the wave function of the $b^3\Sigma^+$ state is mixed into the wave functions of the  $A^1\Pi, v=0$ state; the amount of triplet character is the same for all levels in the $A$ state. The triplet character of the A$^1\Pi, v=0$ state, causes an $A^1\Pi, v=0 \rightarrow a^3\Pi, v=0$ loss below the 10$^{-6}$ level from the main $A^1\Pi - X^1\Sigma^+$ laser cooling transition.\\

\textit{This article has been accepted for publication in Molecular Physics, published by Taylor \& Francis.}
\bibliographystyle{tfo}
\bibliography{AlF}

\end{document}